\documentclass[a4paper,11pt]{article}
\usepackage[top=0.8in, left=0.8in, right=0.8in,bottom=1in]{geometry}
\usepackage{amsfonts, amssymb, latexsym, amsmath, mathtools}
\usepackage[utf8]{inputenc}
\usepackage[T1]{fontenc}
\usepackage{microtype}
\usepackage{graphicx}
\usepackage[dvipsnames]{xcolor}
\usepackage{hyperref}
\usepackage{subcaption}
\usepackage{svg}
\usepackage{cite}
\usepackage{braket}
\usepackage{stmaryrd}
\usepackage{physics}
\usepackage[shortlabels]{enumitem}

\usepackage{algorithm}
\usepackage{algpseudocode}

\newtheorem{definition}{Definition}

\usepackage{enumitem}
\usepackage{booktabs}
\usepackage[htt]{hyphenat} % allow texttt linebreaks
\usepackage{algorithm}
\usepackage{algpseudocode}

\usepackage{multicol}

\renewenvironment{abstract}
 {\small
  \begin{center}
  \bfseries \abstractname\vspace{-.5em}\vspace{0pt}
  \end{center}
  \list{}{
    \setlength{\leftmargin}{.5cm}%
    \setlength{\rightmargin}{\leftmargin}%
  }%
  \item\relax}
 {\endlist}

\newcommand\doubleplus{+\kern-1.3ex+\kern0.8ex}
\newcommand\doubletimes{*\kern-0.6ex*\kern0.8ex}

\newlength{\rowidth}% row operation width
\AtBeginDocument{\setlength{\rowidth}{14ex}}

\usepackage{listings}

\definecolor{mLightBlue}{HTML}{88C0D0}
\definecolor{mGreen}{HTML}{a3be8c}
\definecolor{mGray}{rgb}{0.5,0.5,0.5}
\definecolor{mPurple}{HTML}{b48ead}
\definecolor{mOrange}{HTML}{d08770}
\definecolor{backgroundColour}{HTML}{eceff4}
\definecolor{mBlue}{HTML}{5e81ac}
\definecolor{mRed}{HTML}{bf616a}

\lstdefinestyle{C++}{
    backgroundcolor=\color{backgroundColour},   
    commentstyle=\color{mGreen},
    keywordstyle=\color{mBlue},
    numberstyle=\tiny\color{mGreen},
    stringstyle=\color{mPurple},
    basicstyle=\ttfamily\footnotesize,
    breakatwhitespace=false,         
    breaklines=true,                 
    captionpos=b,                    
    keepspaces=true,                 
    numbers=left,                    
    numbersep=5pt,                  
    showspaces=false,                
    showstringspaces=false,
    showtabs=false,                  
    tabsize=2,
    language=C++,
    morekeywords={OP, ptr, __syncthreads, uint32_t, blockDim, threadIdx},
    moredelim=**[is][\color{mOrange}]{@}{@},
    moredelim=**[is][\color{mPurple}]{¤}{¤},
}

\lstdefinestyle{futhark}{
    backgroundcolor=\color{backgroundColour},   
    commentstyle=\color{mGreen},
    keywordstyle=\color{mBlue},
    numberstyle=\tiny\color{mGreen},
    stringstyle=\color{mPurple},
    basicstyle=\ttfamily\footnotesize,
    breakatwhitespace=false,         
    breaklines=true,                 
    captionpos=b,                    
    keepspaces=true,                 
    numbers=left,                    
    numbersep=5pt,                  
    showspaces=false,                
    showstringspaces=false,
    showtabs=false,                  
    tabsize=2,
    morekeywords={let, in, scan, map, map2, reduce},
    moredelim=**[is][\color{mOrange}]{@}{@},
    moredelim=**[is][\color{mPurple}]{¤}{¤},
}

\begin{document}

\title{Approximating Entanglement Based on Abstract Interpretation}
\author{
  Aske Nord Raahauge\\
  \texttt{aske.n.r@di.ku.dk}
  \and
  Martin Bom Marchioro\\
  \texttt{cjr485@alumni.ku.dk}\\
  % [0.5cm]{\small Martin Elsman} \\
  % \small \texttt{mael@di.ku.dk}
  \and
  Rasmus Ross Nylandsted\\
  \texttt{kpn134@alumni.ku.dk}
}
\date{\today}

\maketitle

\begin{multicols}{2}

\begin{abstract}
Entanglement is a fundamental property of quantum systems, essential for non-trivial quantum programs. Identifying when qubits become entangled is critical for circuit optimization, and for arguing for the correctness of quantum algorithms. This paper presents a static analysis method for approximating entanglement by extending an already existing abstract interpretation, thus avoiding the exponential slowdown of an exact analysis. The approach is shown to be sound and an implementation is provided in Standard ML with linear-time scalability. 
\end{abstract}

\section{Introduction}
Quantum circuits enable the manipulation of qubits to perform operations that exploit quantum mechanical principles such as superposition and entanglement. Among these, entanglement — a unique quantum property that has no classical counterpart — plays a pivotal role in non-trivial quantum algorithms. Understanding when and how qubits become entangled within a circuit is crucial for advancing various quantum technologies and optimizing their performance. 

Identifying instances of entanglement formation in quantum circuits has profound implications across several domains, e.g., circuit optimization and error detection, where knowing when qubits are entangled can help optimize gate arrangements and aid in error mitigation by reducing noise in quantum systems.

The problem of determining when a quantum system is in an entangled state is dubbed \textit{the separability problem}, and has previously been shown to be NP-hard\cite{nphard}. Instead of an exact analysis, approximation schemes have been proposed to accurately approximate the separability of a quantum state in polynomial time with respect to the number of qubits.

We attempt to develop a tool for static analysis of the entanglement of quantum variables, representing qubits in a circuit. Our work represents an implementation of the abstract domain introduced by Perdrix \cite{perdrix} with the expanded notion of levels as presented by Assolini et al.\cite{assolini}. Our intention is to expand the existing quantum computation modeling framework developed for the Advanced Topics in Programming Languages course in Standard ML\cite{sml-quantum}, by incorporating this approximate static entanglement analysis.

\paragraph{Novel Contributions}

\begin{itemize}
    \itemsep-0.2em
    \item A formal adaptation of the \texttt{QIL}-based analysis; conforming to the SML Framework \cite{perdrix2}.
    \item A formal definition of qubit levels, and a derived ruleset for managing such levels \cite{assolini}.
    \item A thorough description of soundness for the adapted analysis.
    \item An implementation of the analysis.
\end{itemize}

\paragraph{Code execution} 
The analysis implementation can be found within the directory: \\ \texttt{atpl-sml-quantum/entanglement\_analysis}. The main algorithm is implemented in \\ \texttt{analysis\_levels.sml}, while auxiliary functions can be found in the file \texttt{helpers.sml}. Lastly, tests are located in \texttt{test.sml}. \medskip

To run the analysis (with the example provided in Assolini et al \cite{assolini}), simply run 

\begin{verbatim}
    $ make
\end{verbatim}

And similarly to run the corresponding tests:

\begin{verbatim}
    $ make test
\end{verbatim}

\section{Background}
We recall the basis for understanding entanglement of quantum variables as used by Perdrix\cite{perdrix, perdrix2} and Assolini et al.\cite{assolini}. We assume that the reader has the academic qualifications corresponding to an introductory course in quantum computation. We will refer to the circuit model for quantum computation, where wires represent individual qubits, which are abbreviated as quantum variables when used in the abstract domain.

For denoting quantum states we use ket notation where we denote a linear combination as $\ket{\psi} = \alpha \ket{0} + \beta \ket{1}$ with the probabilities of collapsing to the first and second state being $|\alpha|^2$ and $|\beta|^2$ respectively.

\subsection{Conventions}
Given the ket description of a quantum state, we will use the term \emph{subsystem} when referring to a factor in a tensor product, and \emph{substate} when referring to a term in a summation of kets with non-zero probability/amplitude. For a state composed of $m$ substates, we will refer to $\chi_i^k \in \{0, 1\}$ as the value of the $i$'th qubit in substate $k$. Similarly, we let $\neg \chi_i^k$ be its opposing value. For example, the state: \[\ket{\psi} = \frac{1}{\sqrt{2}} (\ket{00} + \ket{11}) \otimes \ket{00}\] can be decomposed into subsystems $\frac{1}{\sqrt{2}} (\ket{00} + \ket{11})$ and $\ket{00}$, with the first consisting of substates $\frac{1}{\sqrt{2}}\ket{00}$ and $\frac{1}{\sqrt{2}}\ket{11}$, where we have $\chi_{0}^0=0$, $\chi_{1}^0=0$, $\chi_{0}^1=1$ and $\chi_{1}^1=1$. With this, an $n$ qubit state $\ket{\psi}$ composed of $m$ substates with amplitudes $\alpha_k$, could e.g. be written as:
$\ket{\psi} = \sum_{k=1}^{m} \alpha_k \ket{\chi_0^k ... \chi_{n-1}^k}_k$ \\
% --- More precise, but verbose and ugly
% $\neg \chi_i^k \in (\{0, 1\}~ \backslash ~\{\chi_i^k\})$

In all subsequent examples of $n$ qubit states, we will describe the qubits as belonging to a set $Q = \{q_0,...,q_{n-1} \}$, where each qubit $q_i$ is represented with an index corresponding to its position in the ket-based state description. For the $\ket{+}$ and $\ket{-}$ states, we refer to the symbol $\delta \in \{+, -\}$ with $\neg \delta$ as the opposing state: Such that if $\ket{\delta} = \ket{+}$, then $\ket{\neg \delta} = \ket{-}$ and vice versa. To denote for which qubit(s) we apply a gate \texttt{G}, we use subscripts such as $\texttt{G}_{q_i}$ or $\texttt{G}_{q_i,q_j}$, when showcasing computations. Further more, we will represent an arbitrary instance of the set of Pauli gates as $\sigma \in \{X, Y, Z\}$.

\subsection{Language Frameworks}

The entanglement analysis introduced by Perdrix \cite{perdrix} was originally designed for the high-level \texttt{QIL} quantum language \cite{perdrix2}. In \texttt{QIL}, gates are applied to qubits sequentially, while conditionals and loops can be facilitated by conducting intermediary measurements and evaluating the collapsed result (for more details, see \cite{perdrix2}).

In contrast, the SML Framework provides a circuit/state-based syntax, representing the parallel nature of quantum systems, where the order of gates is ambiguous. It additionally provides no support for intermediary measurement. As opposed to Perdrix' analysis (where mixed states can occur as a consequence of measurements) the SML Framework operates on pure states only. For our adaptation of the analysis, we are thus likewise limited to the rule-subset relating to pure states.

Syntactically, parallel updates are written with infix tensor notation (\texttt{**}) between gates; represented as: \texttt{I}, \texttt{H}, \texttt{T}, \texttt{X}, \texttt{Y}, \texttt{Z}, \texttt{SW} (Swap), \texttt{CX} (CNOT), \texttt{CY}, \texttt{CZ} and so on. Since Perdrix' rules are defined for the controlled CNOT only, we subsequently ignore all other control-gates in the adapted analysis.

Gate compositions, i.e. columns in the quantum circuit, are represented with the operator \texttt{oo}. However, to not clash with the $\circ$-notation for function composition, we will henceforth represent this with the symbol \texttt{++} instead.

\subsection{Entanglement}
We say that a quantum state becomes entangled as a result of some computation, when it enters a superposition of strongly correlated substates, in such a way that the states of the individual quantum variable cannot be described independently of the state of the others.

More formally, we follow the definition of entanglement by \cite{assolini} and say that a composite quantum state $|\psi\rangle_{q_1, q_2} \in \mathcal{H}_{q_1} \otimes \mathcal{H}_{q_2}$ is separable if and only if it can be written as $|\psi\rangle_{q_1, q_2} \in |\phi_1\rangle \otimes |\phi_2\rangle$ for some states $|\phi_1\rangle \in \mathcal{H}_{q_1}$ and $|\phi_2\rangle \in \mathcal{H}_{q_2}$. A state $|\psi\rangle_{q_1, q_2}$ is entangled if and only if it is not separable.

\subsection{Standard and Diagonal Basis}
We recollect the definition of bases as introduced by Perdrix\cite{perdrix}. Intuitively we say that a given quantum variable $q$ is in the standard basis if for $\ket{\psi}_{q} = \alpha\ket{0} + \beta\ket{1}$ it is given that $\alpha = 0 \lor \beta = 0$.

Similarly $q$ is in the diagonal basis if it can be represented as $\ket{\psi}_{q} = \alpha\ket{+} + \beta\ket{-}$ and we have that $\alpha = 0 \lor \beta = 0$. Where 

\[
\gamma\ket{\pm} = \gamma\left( \frac{\ket{0} \pm \ket{1}}{\sqrt{2}} \right)
\]

More formally the basis labeling function $b^{q \mapsto l}$ is defined as the following mapping: \\ $B^Q \, : \, Q \rightarrow \{\textbf{s}, \textbf{d}, \top \}$, which is defined for any $q \in Q$ as follows 
\[
b^q = 
\begin{cases} 
  % \bot & \textit{if q is in both standard and diagonal basis in } \psi \\
  \mathbf{s} & \textit{if q is only in the standard basis in } \psi  \\
  \mathbf{d} & \textit{if q is only in the diagonal basis in } \psi  \\
  \top & \textit{otherwise} 
\end{cases}
\]

We explicitly disregard the case where a quantum variable exists in both the standard and diagonal basis at the same time as denoted by the bottom label $(\bot)$ by Perdrix\cite{perdrix}. As stated, this is possible because our investigation builds on top of the expanded domain by Assolini et al.\cite{assolini}, which is not defined for mixed states. In general, it is impossible for a pure state to be in both the standard and diagonal basis. 
% For a definition of bases for mixed states, we refer to Perdrix' paper \cite{perdrix}.

\subsection{Levels}
The notion of levels, as introduced by Assolini et al.\cite{assolini}, is a special kind of entanglement defined as follows:

\begin{definition}{}
Given two variables $a$ and $b$ in a state $\ket{\psi}$, we say that two entangled variables are \textit{on the same level} if, by measuring one of them, the other also collapses to a base state.
\end{definition}

Where a base state is referring to one of the orthonormal vectors in the Hilbert space that spans the space of possible quantum states for a system. I.e. for the Hilbert space of dimension 1 the basis states are $\ket{0}$ and $\ket{1}$.

To formalize our discussion of levels, we provide a novel algebraic definition to help identify when \emph{levelness} occurs in a quantum state:

\begin{definition}{}
Given an $n$ qubit quantum state $\ket{\psi}$ with $m$ substates: 

\[\ket{\psi} = \sum_{k=1}^m \alpha_k \cdot 
\ket{\chi_0^k ... \chi_{n-1}^k}_k\]

We say that two qubits $q_i$ and $q_j$ are on the same level, if and only if:

\begin{enumerate}[1)]
    \item Both $q_i$ and $q_j$ are in a superposition.
    
    \item across all $m$ substates, it is either the case that $\chi_i^k = \chi_j^k$ or $\chi_i^k =~ \neg \chi_j^k$.
\end{enumerate}
\end{definition}{}

\paragraph{Intuition}
From the definition, property \emph{1)} is necessary, seeing as entanglement implies superposition. If one qubit was not in a superposition, its bit value would be the same across all substates; thus allowing one to tensor it out from the remaining subsystem.

For qubits satisfying property \emph{1)}, property \emph{2)} intuitively states that the substate value of qubit $q_i$ is always dependent on $q_j$ and vice versa. Together, these two properties imply general entanglement (a \emph{weaker} concept), as one cannot tensor out a superposition qubit, whose substate value directly depends on another such qubit.

To illustrate this, consider the following 2-qubit state examples, where the first and second qubits are on the same and different levels respectively:

\begin{itemize}
    \item \emph{Same Level} \\
    $\ket{\psi_1} = \frac{1}{\sqrt{2}} (\ket{00} + \ket{11})$, \\
    $\ket{\psi_2} = \frac{1}{\sqrt{2}} (\ket{10} + \ket{01})$ 
    \item \emph{Different Level} \\
    $\ket{\psi_3} = \ket{00}$, \\
    $\ket{\psi_4} = \frac{1}{\sqrt{3}} (\ket{00} + \ket{01} + \ket{11})$
\end{itemize}

As seen for $\ket{\psi_1}$ and $\ket{\psi_2}$, both qubits are on the same level and cannot possibly be tensored out. By collapsing one of them, you will know the base state of the other.

For $\ket{\psi_3}$, the qubits are on a different level, given that none of them are in a superposition. For $\ket{\psi_4}$, both qubits are in a superposition (and entangled), but do not depend on one another across all substates. Specifically, the substates '$\ket{00}$' and '$\ket{01}$' violate the decree of property \emph{2)}: That qubit values be either different or the same across \emph{all} substates. That is, measuring $q_1$ as 0 collapses the state into the superposition $\ket{+}$, rather than one of the base states $\ket{0}$ or $\ket{1}$.

\paragraph{Transitivity}
The notion of levels is transitive (as shown by Assolini et al. \cite{assolini}), such that a qubit $q_i$ on the same level as $q_j$, and $q_j$ on the same level as $q_z$, implies that $q_i$ is also on the same level as $q_z$. This arises from the fact that if $q_j$ is dependent on $q_i$, and $q_z$ dependent on $q_j$, then $q_z$ must be dependent on $q_i$ (and vice versa) thus satisfying \emph{property 2)}. Seeing as all of $q_i, q_j, q_z$ are leveled with some other qubit, $q_i$ and $q_z$ must additionally be in a superposition, thus satisfying property \emph{1)}.

For a collection of qubits to be on the same level, it intuitively means that between any two substates, the leveled qubits are either all the same or all different, when comparing them pairwise across substates.

\paragraph{Application}
If \emph{levelness} between qubits is known, it can be used to identify some specific situations, where these may become \emph{disentangled} once again, thereby improving the entanglement approximation analysis. Specifically, If two qubits are entangled on the same level, the application of a \texttt{CNOT}/\texttt{CX} gate between them will disentangle the target qubit from all other qubits in the remaining subsystem.

This behavior is a direct result of the dependency between leveled qubits: By applying \texttt{CNOT} control to one of the qubits in a substate, the other only flip if the first has value 1. Since the second qubit has a different value depending on whether the first is 0 or 1 in a substate, the combination that flips will cause the second qubit to be the same as the value in the other (unflipped) combination. As the second qubit thus acquires the same value for all substates, it can be tensored out, and thereby shown to be disentangled from the remaining subsystem.

As an example, observe the same-level states $\ket{\psi_1}$ and $\ket{\psi_2}$ after a \texttt{CNOT} has been applied with the first qubit as control and the second as target:
\footnotesize
\begin{align*}
    CX_{q_0,q_1} \ket{\psi_1} &= CX_{q_0,q_1} (\frac{1}{\sqrt{2}} (\ket{00} + \ket{11})) \\
    &= \frac{1}{\sqrt{2}} (\ket{00} + \ket{10}) \\
    &= \frac{1}{\sqrt{2}} ((\ket{0} + \ket{1}) \otimes \ket{0})
\end{align*} 

\begin{align*}
    CX_{q_0,q_1} \ket{\psi_2} &= CX_{q_0,q_1} (\frac{1}{\sqrt{2}} (\ket{10} + \ket{01})) \\
    &= \frac{1}{\sqrt{2}} (\ket{11} + \ket{01}) \\
    &= \frac{1}{\sqrt{2}} ((\ket{1} + \ket{0}) \otimes \ket{1})
\end{align*}
\normalsize

\section{Extended Abstract Semantics}
Given that we represent our systems as quantum circuits, the existing syntax as defined by Perdrix\cite{perdrix} is ambiguous with respect to the order of execution when quantum gates are introduced in parallel (I.e. as tensor products separated by compositions). In the case of the quantum circuit simulation framework in Standard ML, this occurs whenever quantum gates are separated with the \texttt{**} operator. We now need to extend the denotational abstract semantics\cite{perdrix} to account for when gates are placed in parallel. This new operation should bind harder than sequential commands (defined by Perdrix as $\llbracket C_1; C_2 \rrbracket^\natural (b, \pi)$), meaning that parallel operations are grouped over sequential ones. \medskip

Additionally, in order to incorporate the notion of being on the same level, we define an extended abstract domain. 

\begin{definition}{}
We define the abstract domain $\mathcal{B}^Q = B^Q \times \Pi_\varsigma^Q \times \Pi_\iota^Q$, where \\ $B^Q = Q \rightarrow \{\mathbf{s}, \mathbf{d}, \top\}$ is the basis label function. $\Pi_\varsigma^Q$ is a collection of subsets of non-separable qubits, and $\Pi_\iota^Q$ is a collection subsets of qubits on the same level. Thus given a set of quantum variables $Q$, we define an abstract state as
\[
s^Q = \left\{ (b, \pi_\varsigma, \pi_\iota) \; | \; b \in B^Q, \; \pi_\varsigma \in \Pi_\varsigma^Q, \; \pi_\iota \in \Pi_\iota^Q  \right\}
\]
\end{definition}

As an example, a state with $n = 4$ qubits, where:
\begin{align*}
    &\pi_\varsigma = \{\{q_0, q_3\}, \{q_1, q_2\}\},~~ \text{and} \\
    &\pi_\iota = \{\{q_0, q_3\}, \{q_1\}, \{q_2\}\}
\end{align*} 

would mean that qubit $q_0$ and $q_3$ are approximated to be entangled, as well as qubit $q_1$ and $q_2$ as seen from $\pi_\varsigma$. $\pi_\iota$ then states that $q_0$ and $q_3$ are additionally on the same level, while $q_1$ and $q_2$ are not.

As stated, the analysis proposed by Perdrix is approximative, meaning that we may over-approximate the entanglement/non-separability of variables. As a result, this could cause separable qubits to be incorrectly identified as 'entangled'.

When extending the abstract semantics to include gates being run in parallel we further more need information on the numbering of quantum variables in the set. In order to formally define this we introduce the notion of \textit{height} as follows

\begin{definition}{}
Given a circuit $\langle C, Q \rangle$ we say that its height is defined on the command $C$ as follows

\begin{equation*}
\begin{split}
\text{height}(C) : \; & \texttt{I}, \; \sigma, \; \texttt{H}, \; \texttt{T}  \mapsto 1 \\
    & | \; \texttt{CX}, \; \texttt{SW} \mapsto 2 \\
    & | \; C_1 \doubleplus C_2 \mapsto \text{height}(C_1) \\
    & | \; C_1 \doubletimes C_2 \mapsto \text{height}(C_1) + \text{height}(C_2)
\end{split}
\end{equation*}

\end{definition} ~

Additionally we define the usual notion of substitution as $(b, \pi_\varsigma, \pi_\iota)[q_i \rightarrow q_j]$; meaning that each reference to $q_i$ is replaced with $q_j$ for all state components. Borrowing notation from Perdrix \cite{perdrix}, we also let $\pi \vee [q_i, q_j]$ denote an updated partition, where the set containing $q_i$ is united with the set containing $q_j$ in $\pi$. With '$-$' denoting set difference, we formally define such an update $\pi' = \pi \vee [q_i, q_j]$ as follows:
\begin{equation*}
\begin{split}
\pi' = ((\pi \; - \; \{ \pi_i^* & \}) \; - \; \{ \pi_j^* \})) ~\cup~ \{ \pi_i^* ~\cup~ \pi_j* \} \\
&\text{where } \pi_i* \in \pi \text{, such that } q_i \in \pi_i^*, \\
&\text{and } \pi_j* \in \pi \text{, such that } q_j \in \pi_j^*
\end{split}
\end{equation*}

Finally we let the ``split'' operation $\pi \backslash q$ denote the updated partition, where $q$ is removed from its previous set and placed in its own singleton set. As above, we formally define the update $\pi' = \pi \backslash q$ as: \\
\begin{equation*} % split
\begin{split}
\pi' = ( \pi \; - \; \{ \pi^* & \} ) \; \cup  \; \{ \pi^* \; - \; q , \{ q \} \} \\
& \text{where } \pi^* \in \pi \text{ such that } q \in \pi^*
\end{split}
\end{equation*} \\

\subsection{Analysis Overview}

With this, we can now present our extended analysis rules, with its own denotational semantics for a given circuit $C$, as presented below. For ease of notation, we were advised to present the set of qubits as integers, such that a qubit $q_i$ corresponds to the qubit at index $i$, whilst $q_{i+x}$ corresponds to the qubit at index $i+x$. An alternate version of the semantics, with qubits modelled as the usual abstract entities, is presented in Appendix \ref{sec:app2}.

\end{multicols} ~

% SEMANTICS
\begin{definition}{}\label{def:semantics}
For any program $\langle C, Q \rangle$, let $\llbracket C \rrbracket^\natural_q \, : \, \mathcal{B}^Q \rightarrow \mathcal{B}^Q$ be defined as follows for any \\ $(b, \pi_\varsigma, \pi_\iota) \in \mathcal{B}^Q$

\begin{equation*}
\begin{split}
\llbracket \texttt{I} \rrbracket^\natural_q (b, \pi_\varsigma, \pi_\iota) &= (b, \pi_\varsigma, \pi_\iota) \\
\\
\llbracket C_1 \doubletimes C_2 \rrbracket^\natural_q (b, \pi_\varsigma, \pi_\iota) &= \llbracket C_2 \rrbracket^\natural_{q'} \circ \llbracket C_1 \rrbracket^\natural_q (b, \pi_\varsigma, \pi_\iota) \textit{ where } q' = q + \text{height}(C_1) \\
\\
\llbracket C_1 \doubleplus C_2 \rrbracket^\natural_q (b, \pi_\varsigma, \pi_\iota) &= \llbracket C_2 \rrbracket^\natural_q \circ \llbracket C_1 \rrbracket^\natural_q (b, \pi_\varsigma, \pi_\iota) \\
\\
\llbracket \sigma \rrbracket^\natural_q (b, \pi_\varsigma, \pi_\iota) &= (b, \pi_\varsigma, \pi_\iota) \\
\\
\llbracket \texttt{H} \rrbracket^\natural_q (b, \pi_\varsigma, \pi_\iota) &= (b^{q \mapsto \textbf{d}}, \pi_\varsigma, \pi_\iota) \textit{ if } b^q = \textbf{s} \\
&= (b^{q \mapsto \textbf{s}}, \pi_\varsigma, \pi_\iota) \textit{ if } b^q = \textbf{d} \\
&= (b, \pi_\varsigma, \pi_\iota \backslash q) \textit{ otherwise} \\
\\
\llbracket \texttt{T} \rrbracket^\natural_q (b, \pi_\varsigma, \pi_\iota) &= (b^{q \mapsto \top}, \pi_\varsigma, \pi_\iota) \textit{ if } b^q = \textbf{d} \\
&= (b, \pi_\varsigma, \pi_\iota) \textit{ otherwise} \\
\\
\llbracket \texttt{CX} \rrbracket^\natural_q (b, \pi_\varsigma, \pi_\iota) &= (b, \pi_\varsigma, \pi_\iota) \textit{ if } b^q = \mathbf{s} \textit{ or } b^{q+1} = \mathbf{d} \\
&= (b^{q, q+1 \mapsto \top}, \pi_\varsigma \vee [q, q+1], \pi_\iota \vee [q, q+1]) \textit{ if } b^q = \mathbf{d} \textit{ and } b^{q+1} = \mathbf{s} \\
&= (b^{q+1 \mapsto \mathbf{s}}, \pi_\varsigma \backslash q+1, \pi_\iota \backslash q+1) \textit{ if } \exists \mathcal{P} \in \pi_\iota, \textit{ where } \{q, q+1\} \subseteq \mathcal{P}  \\
&= (b^{q, q+1 \mapsto \top}, \pi_\varsigma \vee [q, q+1], \pi_\iota) \; \textit{otherwise} \\
\\
\llbracket \texttt{SW} \rrbracket^\natural_q (b, \pi_\varsigma, \pi_\iota) &= (b, \pi_\varsigma, \pi_\iota)[q \mapsto q+1, q+1 \mapsto q]
\end{split}
\end{equation*}

\end{definition}
~\vspace{80pt}

\begin{multicols*}{2}

Given an n-qubit system, the initial state $\ket{00\dots0}$ of our abstract domain is represented as $(b, \pi_\varsigma, \pi_\iota) \in \mathcal{B}^Q$, where:
\[b=\{q_0\mapsto\textbf{s},q_1 \mapsto\textbf{s},\dots,q_n \mapsto\textbf{s}\}\]
\[\pi_\varsigma=\{\{q_0\},\{q_1\},\dots,\{q_n\}\}\]
\[\pi_\iota=\{\{q_0\},\{q_1\},\dots,\{q_n\}\}\]
In other words, all basis labels are set to the standard basis \textbf{s}, while all qubits are initially marked as separable and on their own level. \\

Intuitively, the quantum operations manipulate our abstract state as follows:
\begin{itemize}
    \item Both the tensor (\texttt{**}) and sequence ($++$) operators work as function composition, and start by evaluating the leftmost circuit, then applying this result to the rightmost. The only difference being that \texttt{**} updates $q$ according to the height of the evaluated sub-circuit. 
    \item All single-qubit gates except Hadamard (\texttt{H}) do not modify entanglement nor levels, i.e. the  entanglement ($\pi_\varsigma$) and levels ($\pi_\iota$) sets remain unchanged. Any pauli operator and the identity gate preserve the standard or diagonal basis of the qubit, while the \texttt{T} gate only preserves the standard basis. Lastly, the Hadamard gate flips the standard and diagonal basis, while also splitting the qubit from its set in $\pi_\iota$.
    \item \texttt{CX} and \texttt{SW} work by applying the gates to qubit $q$ and $q+1$. Note that this differs from Perdrix' semantics, where \texttt{CX} can be applied to any 2 arbitrary qubits $q_1$ and $q_2$. However, this does not make our analysis less potent, as one can swap the qubits into any combination before performing the controlled not gate. \texttt{SW} works by swapping the values of $q$ and $q+1$ in every set of our state. \texttt{CX} is a bit more involved, and will sometimes join and split entanglement and levels. If the control qubit is in the standard basis, or if the target qubit is in the diagonal basis, then the state is preserved. If $b^q=\textbf{d}$ and $b^{q+1}=\textbf{s}$ then we join $q$ and $q+1$ sets in both $\pi_\varsigma$ and $\pi_\iota$, and change their basis labels to $\top$. If both control and target qubits reside in the same levels subset, then we split the target qubit from $\pi_\varsigma$ and $\pi_\iota$ and set its state to \textbf{s}. Finally, if none of the above are applicable, we join the entanglement subsets for the qubits and change both their labels to $\top$.   
\end{itemize}

The intuition behind these updates will be discussed in the subsequent sections.

\section{Mechanics \& Soundness of Analysis}

We now outline the mechanics and soundness of each component of the analysis: Namely in terms of its general structure, label updates, and separability/level maintenance. Although a completely formal description would be desired, we deem such discussions to be outside the scope of this project. 

% TODO: meget eksplicit skriv at CX og H regler er inferred og vores egne

\subsection{General Structure}
To avoid name-clashes, we will refer to $i$ as the circuit height in the following subsection.

\paragraph{Termination} Seeing as there are no loops within the SML framework, and only a constant amount of gates for each circuit, the analysis' approach of iteratively processing each gate is guaranteed to terminate.

\paragraph{Qubit Updates} In the syntax tree of an SML circuit, both composition and tensor nodes will prompt our analysis to first traverse its left-most sub-tree. Initially we will therefore guarantee to always analyze the first gate of the first qubit $q_0$. When initializing the analysis with height $i = 0$, it will propagate to this gate, and ensure that it is the state of the first qubit (and optionally the second for 2-qubit gates) that is updated when analyzing the gate. A 1-qubit gate then adds $1$ to this height, while a 2-qubit gate adds $2$. At that point $i$ will therefore correspond to the index of the next qubit to be updated.

After this ``base case'' the analysis might in general return to a tensor ($**$) or a composition ($++$) parent node. For a tensor, the updated height from the left child is propagated when analyzing the right child, which thereby supplies the correct index for the next qubit to have its gate analyzed.

If the analysis returns to a composition node, we know that a well-formed circuit will update the same span of qubits in its left and right sub-tree. The updated height of the first sub-tree is therefore ignored, in favor of the old height, which ``resets'' the analysis for the same span of qubits, when analyzing the right sub-tree of the composition. The updated height of the right sub-tree is thereafter returned to the parent node to properly accommodate any tensor nodes higher up in the syntax tree.

\paragraph{Parallel ordering}
Unlike the sequential \texttt{QIL} language of Perdrix' analysis \cite{perdrix} \cite{perdrix2}, the circuit representation of the SML framework allows for a parallel representation of quantum gates. When adapting the analysis, one must therefore ensure that the relative order, in which we analyze parallel gates does not affect the final result of the analysis.

However, since each gate can only update the qubit state at the corresponding height (with 2-qubit gates ranging over 2 qubits), the locally defined label-mapping of qubits will be the same regardless of the analysis order.

For separability and level partitions (which range over global sets), the order of unions between sets is likewise irrelevant, seeing as set union is commutative.

For ``splitting'' a set, the analysis is capable of removing a single qubit $q$ from the given set, and turn it into its own singleton set. Since the corresponding (\texttt{CX}) split-operation, will be the only gate acting on $q$ in the parallel order; no other operation will be dependent on whether $q$ has been split or not. In particular, if there is some operation that joins the set of $q$ with some other set (prior to splitting $q$), the join must necessarily be facilitated by some other qubit in the subset of $q$ -- it is therefore irrelevant whether $q$ is part of, or have already been split from such a set. As such, splitting cannot disturb the commutativity of set joins. Dually, since a split variable is a singleton, which cannot be referenced by any other operation, its state cannot be further changed after the split, while also being unaffected by any partition changes before the split.

In conclusion, the order in which we analyze parallel operations is therefore irrelevant to the result of the analysis.

\subsection{Labels}\label{sec:labels}

We now describe the mechanical soundness of re-labelling qubits throughout the analysis; particularly with regards to the \texttt{s} and \texttt{d} labels (seeing as $\top$ can freely refer to any basis).

As stated, a qubit basis is independent of any global amplitude of the state. A qubit will be in the standard basis, if it is not in superposition; i.e. if it corresponds to a state $\ket{\chi}$ for some $\chi$ (i.e. $\ket{0}$ or $\ket{1}$). It will be in the diagonal basis if corresponding to a state $\ket{\delta}$ for some $\delta$ (i.e. $\ket{+}$ or $\ket{-}$). We write '$\pm$', if the state amplitude could potentially have either sign (although not all state/sign-combinations may be valid).

At their starting point, each qubit is assumed to be in the standard $\ket{0}$ state, and is therefore initially labeled \texttt{s}.

\paragraph{Identity \& Pauli-Gates}
Both the identity \texttt{I} as well as the Pauli-gates \texttt{X}, \texttt{Y}, and \texttt{Z} preserve the basis of their qubits as seen:

\begin{multicols}{2}
\small
\textbf{Standard basis:} ~\vspace{2pt}
\begin{itemize}
    \scriptsize
    \itemsep0em
    \item $\texttt{I} \ket{\chi} = \ket{\chi}$
    \item $\texttt{X} \ket{\chi} = \ket{\neg \chi}$
    \item $\texttt{Y} \ket{\chi} = \pm i \ket{\neg \chi}$
    \item $\texttt{Z} \ket{\chi} = \pm \ket{\chi}$
    \normalsize
\end{itemize}

\columnbreak

\small
\textbf{Diagonal basis:}
\begin{itemize}
    \scriptsize
    \itemsep0em
    \item $\texttt{I} \ket{\delta} = \ket{\delta}$
    \item $\texttt{X} \ket{\delta} = \pm \ket{\delta}$
    \item $\texttt{Y} \ket{\delta} = \pm i \ket{\neg \delta}$
    \item $\texttt{Z} \ket{\delta} = \ket{\neg \delta}$
    \normalsize
\end{itemize}
\end{multicols}

\paragraph{Hadamard \& T-gate}
Hadamard switches between \texttt{s} and \texttt{d}, while the T-gate preserves \texttt{s}:

\begin{multicols}{2}
\small
\textbf{Hadamard}: \vspace{2pt}
\begin{itemize}
    \scriptsize
    \itemsep0em
    \item $\texttt{H} \ket{\chi} = \ket{\delta}$
    \item $\texttt{H} \ket{\delta} = \ket{\chi}$
    \normalsize
\end{itemize}

\columnbreak

\small
\textbf{T-gate}:
\begin{itemize}
    \scriptsize
    \itemsep0em
    \item $\texttt{T} \ket{0} = \ket{0}$
    \item $\texttt{T} \ket{1} = e^{i\pi/4} \ket{1}$
\end{itemize}
\end{multicols}
\normalsize

However, given that $\texttt{T} \ket{+} = \frac{1}{\sqrt{2}} (\ket{0} +  e^{i\pi/4} \ket{1})$, the T-gate does not preserve the diagonal basis, and the label is therefore set to $\top$ in this case.

\paragraph{CNOT}

The \texttt{CX} gate preserves the \textbf{s}/\textbf{d} bases of qubits, if the control $q_i$ is in \texttt{s} basis or target $q_j$ in \texttt{d} basis.

In this scenario; $q_i$ and $q_j$ cannot be entangled, as in the resulting state, at least one of them can be described as an isolated subsystem (i.e. a basis-vector of either the standard or diagonal basis). If one of them is a non-entangled $\top$-labeled qubit, it may either be in a base state, and thus be equivalent to an \textbf{s}-labelled qubit, or in a general superposition: $\ket{\psi} = \alpha_0 \ket{0} + \alpha_1 \ket{1}$.

With this in mind, a detailed analysis of each case is provided below, where we consider only the isolated subsystems corresponding to the two qubits (which we assume to be non-entangled in general). As can be verified, the bases are preserved in each case, except for when a $\top$-labeled qubit is in general superposition. However, seeing as the $\top$ label may refer to any basis, it remains accurate even if the exact basis is not preserved. 

The result also generalizes to $\top$-labeled qubits, which \emph{are} entangled, seeing as they display the same underlying behavior.

\small
\begin{itemize}
    \item \textbf{Control s, Target s} \\
        \scriptsize
        $\texttt{CX}_{q_i, q_j} \ket{0} \otimes \ket{\chi_j} = \ket{0} \otimes \ket{\chi_j}$ \\
        $\texttt{CX}_{q_i, q_j} \ket{1} \otimes \ket{\chi_j} = \ket{1} \otimes \ket{\neg \chi_j}$
        \small

    \item \textbf{Control s, Target d} \\
        \scriptsize
        $\texttt{CX}_{q_i, q_j} \ket{0} \otimes \ket{\delta} = \ket{0} \otimes \ket{\delta}$, \\
        $\texttt{CX}_{q_i, q_j} \ket{1} \otimes \ket{\delta} = \ket{1} \otimes \pm \ket{\delta}$
        \small

    \item \textbf{Control s, Target $\top$} \\
        \scriptsize
        $\texttt{CX}_{q_i, q_j} \ket{0} \otimes (\alpha_0 \ket{0} + \alpha_1 \ket{1})
        = \ket{0} \otimes (\alpha_0 \ket{0} + \alpha_1 \ket{1})$, \\
        $\texttt{CX}_{q_i, q_j} \ket{1} \otimes (\alpha_0 \ket{0} + \alpha_1 \ket{1})
        = \ket{1} \otimes (\alpha_0 \ket{1} + \alpha_1 \ket{0})$
        \small

    \item \textbf{Control d, Target d} \\
        \scriptsize
        $\texttt{CX}_{q_i, q_j} \ket{\delta} \otimes \ket{+} = \ket{\delta} \otimes \ket{+}$, \\
        $\texttt{CX}_{q_i, q_j} \ket{\delta} \otimes \ket{-} = \ket{\neg \delta} \otimes \ket{-}$
        \small

    \item \textbf{Control $\top$, Target d} \\
        \scriptsize
        $\texttt{CX}_{q_i, q_j} (\alpha_0 \ket{0} + \alpha_1 \ket{1}) \otimes \ket{+}
            = (\alpha_0 \ket{0} + \alpha_1 \ket{1}) \otimes \ket{+}$, \\
        $\texttt{CX}_{q_i, q_j} (\alpha_0 \ket{0} + \alpha_1 \ket{1}) \otimes \ket{-}
            = (\alpha_0 \ket{0} - \alpha_1 \ket{1}) \otimes \ket{-}$
        \small
\end{itemize}
\normalsize

However, for other cases such as: 
\\ $\texttt{CX}_{q_i,q_j} \frac{1}{\sqrt{2}} (\ket{0} + 1) \otimes \ket{0} = \frac{1}{\sqrt{2}} (\ket{00} + \ket{11})$, \\
the two qubits can become entangled and will therefore not be expressible in the standard nor the diagonal basis. Qubits in any other case will therefore have to be labelled $\top$.

\subsection{Seperability \& Levels}

CNOT is the only operation, wherein two qubits interact and thus have the potential to become entangled. Thus, it is only through the \texttt{CX} gate in the SML framework that two qubits may be joined in the $\pi_\varsigma$ set. Since we always want to over-approximate the non-separability of qubits, the analysis will generally be unsound, if it fails to group $q$ in $\pi_\varsigma$ with a set of qubits, which turn out to be non-separable in reality.

\paragraph{Preservation of Levels}

Given two qubits $q_i$ and $q_j$ on the same level, the effects of the Identity, Pauli, and T-gates will not affect their levelness. The reason lies in the gates' transformation properties, which can be summarized as either: Leaving the state unchanged, changing the amplitudes/phases of substates, or flipping a bit across all substates.

As seen in our earlier definition, the levelness between qubits is solely determined by their bit combination across substates, and is therefore unaffected by amplitude changes. 

For flip operations, the argument qubit is flipped across \emph{all} substates, meaning that if $q_i$ and $q_j$ were on the same level as $\chi^k_i = \chi_j^k$, it will thereafter hold that $\chi^k_i = \neg \chi_j^k$ and vice versa for all substates $k$. Qubit $q_i$ and $q_j$ will therefore still fulfill our definition of levelness even after the transformation.

\paragraph{Destruction of Levels/Entanglement}

As described in the \emph{Levels} subsection, a $\texttt{CX}_{q_i, q_j}$ gate applied on two entangled and leveled qubits, $q_i$ and $q_j$, will disentangle and de-level the target qubit to a base state, which can therefore be granted the \textbf{s} label during analysis.

Additionally however, the application of a Hadamard to either $q_i$ or $q_j$ can be shown to also de-level the qubits. Consider e.g. the following example, where a Hadamard is applied to the first qubit of the 2-qubit bell state $\ket{\psi_1}$ (where qubits are initially entangled on the same level): \small
\begin{align*}
    \texttt{H}_{q_0} \ket{\psi_1} &= \texttt{H}_{q_0} \frac{1}{\sqrt{2}} (\ket{00} + \ket{11}) \\
    &=~ \frac{1}{2} (\ket{00} + \ket{10} + \ket{01} - \ket{11})
\end{align*}
\normalsize

Now, given the presence of e.g. substates '$\ket{00}$' and '$\ket{10}$', we see that qubits are no longer on the same level, but still entangled. As a consequence of such occurrences, we have therefore extended the ruleset from Assolini et al.\cite{assolini} \cite{assolini}, by adding the rule that a leveled qubit will be de-leveled from all others, if acted on by a Hadamard.

\paragraph{Emergence of Levels/Entanglement}
As stated briefly in Section \ref{sec:labels}, a \texttt{CNOT} applied to a control bit outside the \textbf{s} basis, and a target outside the \textbf{d} basis may create entanglement. In this case, seen as it is always safe to over-approximate entanglement, we always group the qubits as non-separable in $\pi_\varsigma$, whenever they are not on the same level. 

The controlled NOT gate is particularly interesting, since it is the only gate that can level two qubits. This occurs when the control qubit is \textbf{d} and the target qubit is \textbf{s}. In this configuration, the resulting entangled state always produces one of the 4 maximally entangled 2-qubit states, thus adhering to the definition of same \emph{lavelness}:
\small
\begin{itemize}
    \item \textbf{Control d, Target s} \\
        \scriptsize
        $\texttt{CX}_{q_i, q_j} \ket{\delta} \otimes \ket{0} = \frac{1}{\sqrt{2}} (\ket{00} \pm \ket{11})$, \\
        $\texttt{CX}_{q_i, q_j} \ket{\delta} \otimes \ket{1} = \frac{1}{\sqrt{2}} (\ket{01} \pm \ket{10})$
        \small

\end{itemize}
\normalsize
Even though that in \cite{assolini} it is claimed that: \emph{"the cx gate makes two variables at the same level if the target is labeled as s"}, we found this not to always be the case. Blindly following this statement can lead to erroneous states, particularly if the control bit is marked as $\mathbf{\top}$. This is illustrated in Figure \ref{xxx}. Step \textbf{5} highlights the first contradiction, where the abstract state marks qubit $q$ and $z$ as being on the same level, while none of the qubits are in an entangled state. This clearly contradicts property \emph{1)} of \textbf{Definition 2}. If we keep applying the shown gates, we eventually reach the final erroneous state: $\frac{1}{\sqrt{2}} (|000\rangle + |110\rangle)_{p,q,z} = \frac{1}{\sqrt{2}} (|00\rangle + |11\rangle)_{p,q} \otimes \ket{0}_z$. where the analysis incorrectly marks qubit $q$ as separable, which is clearly not the case. In fact, it belongs, together with qubit $p$, to a maximally entangled Bell state. \\
As a result, it is only safe to join the levels within a \texttt{CX} gate when control is \textbf{d} and target is \textbf{s}. This implies that we can only always have at most two qubits at the same level in $\pi_\iota$, since the join of levels only can occur with separable qubits as control and target.

\section{Representation in SML}
In order to represent a state $s^Q$ given a set of qubits $Q$ we need to represent the triple $(b, \pi_\varsigma, \pi_\iota)$. Given $s^Q$, the state of the label set $b$ must have exactly one element for each qubit i.e. $|b| = |Q|$. Similarly we note that since the collection $\pi_\varsigma$ and $\pi_\iota$ always store exactly all qubits as parts of their subsets, the total number of individual elements across all subsets within the collections, will always equal the number of qubits $|Q|$.

Knowing this we efficiently represent these sets using three arrays of size $|Q|$. The basis set can be represented as an array of simple enumerations: $\textbf{s}$, $\textbf{d}$ or $\mathbf{\top}$. The collections $\pi_\varsigma$ and $\pi_\iota$ are represented using integer arrays, where the i'th index of an array then represents the i'th qubit in the set. The value at each index $i$ then determines the numbered subset in which the corresponding qubit is in. That is, the array $[\mathbf{0},0,\mathbf{2},2,\mathbf{4}]$ corresponds to the collection $\{\{0,1\},\{2,3\},\{4\}\}$. Note the integers highlighted with bold, correspond to the representative of each subset. The value of the representatives thus always correspond to its index. By including this convention we ensure that each collection maps uniquely to an array state, simplifying the verification of correctness while testing. 

Our analysis function takes a circuit as argument in the form: $c=(H \doubletimes I \doubleplus C X \doubleplus SW \doubleplus C X)$, and then decomposes it to the circuit datatype in the quantum SML framework defined as: ~\bigskip
\small
\begin{verbatim}
datatype t = I | X | Y | Z | H | SW | T
             | Tensor of t * t
             | Seq of t * t
             | C of t
\end{verbatim}
\normalsize
We use \texttt{infix 3 oo} and \texttt{infix 4 **} to denote the precedence relation, meaning that we bind the tensor product harder than sequential operations. Having this representation, we are now able to pattern match the circuit and manipulate our abstract state according to the quantum gates. The functionality has been implemented according to the extended abstract semantics defined in \ref{def:semantics}. The height of the circuit, i.e. the index of the qubit that a given gate needs to manipulate, is passed and updated through the recursive call. The two main operations, join and split, are shown in Appendix \ref{yyy}. Join works by comparing the values of array entries $q$ and $q+1$, i.e. the value of the representatives of the subsets each qubit belongs to, and then replacing all values of the qubits belonging to the same subsets as $q$ or $q+1$ to the smallest of those. By doing this we preserve the representative structure. Split on the other hand, starts by determining if the given qubit is the representative of its subset. If its not it means that its index is unused by other qubits, and thus we can safely change its value to said index. If we want to split the representative of a set, we need to find the next representative, i.e. the lowest index with that value in the array, and change all occurrences with that value. \medskip

We intentionally chose to represent the state using arrays, thus manipulating the state in an imperative fashion. This decision was mainly driven by the fast indexing of elements and to avoid having to create copies of the state with each recursive call. Since we always traverse the circuit tree from left to right, we never need the state of a previous subtree, which otherwise would get overwritten by the most recent gate. Thus, by overwriting the same array at each subsequent gate, causes us to lose information about the state in the previous gates.
However, if future work requires us to keep track of different states at different point in times while traversing the subtrees, the implementation could be modified to adhere to a fully functional approach. This would involve passing the state down the recursive tree alongside the height, while using an immutable data structure such as a list. \medskip

We have also added a less accurate analysis implementation in \textbf{analysis.sml}, which disregards the notion of levels. Comparing its output with the implementation which includes levels, provides insight to the usefulness of levels. This is especially evident when analyzing a circuit consisting of the gates 1-3 in \ref{xxx}. In this case our non-level implementation overapproximates qubit $p$ and $q$ as entangled, while exploiting the extra information provided by the levels, we are able to correctly identify $p$ and $q$ as separable.

\subsection{Runtime complexity}
Representing the states as constant-sized mutable arrays of integers, allows for fast processing of these operations. More specifically, the asymptotic work for both functions, join and split, is simply upper bounded by the array length, determined by the number of qubits. A worst case circuit requires these functions to be called a constant number of times for each gate. This results in the worst case runtime of the algorithm being $O(n \cdot m)$ with $n$ being the number of qubits in the circuit, and $m$ being the total number of gates. This is much superior to other proposed entanglement analysis approaches, especially compared to performing an exact analysis, which operates in exponential time. However, this advantage comes at the cost of precision, as the entanglement can be heavily overapproximated in some cases, especially when analyzing deep circuits.

\section{Future Work}

As previously mentioned, the expressiveness of the SML Framework is limited by its lack of intermediary measurements. As such, it can only represent a subset of general quantum circuits. The full analysis invented by Perdrix \cite{perdrix} \emph{is} however defined for general circuits, where measurements may be used to express conditional and looping constructs. To facilitate this, Perdrix abstracts over density matricies and defines rigid lattices for all components of the domain. The total lattice is shown to be both finite and monotonic, thus demonstrating the feasibility of his framework. 

As future work, the implementation of a full analysis would yield support for general circuits, and additionally provide new opportunities for accurately tracking disentanglement as a consequence of measurements (whereby qubits collapse and thereby become separable).

In our limited analysis of pure states, the only way of approximating disentanglement is by CNOT-flipping leveled qubits, which can only occur under very specific circumstances. Even then however, only one qubit is disentangled, while the other maintains its $\top$ label; thus permanently loosing its ability to later be leveled and thereby disentangled. As a consequence, our approximation will continuously loose information whenever qubits are marked as non-separable.

The \texttt{CX} rule, which entangles and levels, when applied to a \textbf{d} and \texttt{s} labeled qubit, could in principle also work for a $\top$-labeled control qubit in general superposition. Since amplitudes are irrelevant for the definition of levels, the property of having the control be in superposition is sufficient for creating the level-defining bit-combinations, when the target is a base state. As such, if the target qubit is in the standard basis, the only problematic case, is when the control qubit has the $\top$ label, but is a base state in reality. Therefore, if it was possible to enrich the analysis and theory framework with an alternate $\top$ label, where the qubit could be in any basis, \emph{except} the standard basis; one could theoretically improve the accuracy of the analysis, while also allowing more than two qubits to be on the same level (thereby utilizing the transitive property of levelness).

For practical applications however, it would be interesting to benchmark the over-approximation of our analysis with an exact analysis of qubit entanglement, to gain an insight to the degree of overapproximation provided by our analysis.

\section{Conclusion}
In this work, we have introduced an abstract domain based on the concept of separability, and \textit{being on the same level}. Based on Assolini et al.\cite{assolini} we have extended the notion of levels with concrete rule derivations for CNOT and Hadarmard gates. We show how this domain can be used to statically approximate entangled states in quantum systems. Concretizing our work we show how it can be implemented in Standard ML, where we are successfully able to analyze programs in $O(n \cdot m)$ time with respect to the number of qubits and gates in the circuit.

For future reference it would be interesting to extend the existing implementation with conditionals and loop body functionality, in order to measure the performance of approximating entanglement using this approach based on abstract interpretation in comparison to performing an exact analysis.

\end{multicols*}

\newpage
\bibliography{refs}{}
\bibliographystyle{plain}

\newpage
\section{Appendix}

\subsection{Unsafe Leveling Example}

\begin{table}[h]
\[
\begin{array}{llll}
\mathbf{0}: & \{|000\rangle_{p,q,z}\} & \\
            &  \color{blue} \pi_\varsigma=\{\{p\},\{q\},\{z\}\}  & \color{blue} \pi_\iota=\{\{p\},\{q\},\{z\}\} & \color{blue} b=\{p, q, z \rightarrow \mathbf{s} \}   \\
\mathbf{1}: \texttt{h(p)} & \{ \frac{1}{\sqrt{2}} (|000\rangle + |100\rangle)_{p,q,z}\} & \\
            &  \color{blue} \pi_\varsigma=\{\{p\},\{q\},\{z\}\}  & \color{blue} \pi_\iota=\{\{p\},\{q\},\{z\}\} & \color{blue}  b=\{p \rightarrow \mathbf{d}; q, z \rightarrow \mathbf{s} \}   \\
\mathbf{2}: \texttt{cx(p,q)} & \{ \frac{1}{\sqrt{2}} (|000\rangle + |110\rangle)_{p,q,z}\} & \\
            &  \color{blue} \pi_\varsigma=\{\{p,q\},\{z\}\}  & \color{blue} \pi_\iota=\{\{p,q\},\{z\}\} &  \color{blue}  b=\{p, q \rightarrow \top; z \rightarrow \mathbf{s} \}   \\
\mathbf{3}: \texttt{cx(q,p)} & \{ \frac{1}{\sqrt{2}} (|000\rangle + |010\rangle)_{p,q,z}\} & \\
            &  \color{blue} \pi_\varsigma=\{\{p\},\{q\},\{z\}\}  & \color{blue} \pi_\iota=\{\{p\},\{q\},\{z\}\} &  \color{blue}  b=\{q \rightarrow \top; p,z \rightarrow \mathbf{s} \}  \\
\mathbf{4}: \texttt{h(q)} & \{ |000\rangle_{p,q,z}\} & \\
            &  \color{blue} \pi_\varsigma=\{\{p\},\{q\},\{z\}\}  & \color{blue} \pi_\iota=\{\{p\},\{q\},\{z\}\} &  \color{blue}  b=\{q \rightarrow \top; p,z \rightarrow \mathbf{s} \}  \\
\mathbf{5}: \texttt{cx(q,z)} & \{ |000\rangle_{p,q,z}\} & \\
            &  \color{blue} \pi_\varsigma=\{\{p\},\{q,z\}\}  & \color{blue} \pi_\iota=\{\{p\},\{q,z\}\} &  \color{blue}  b=\{q, z \rightarrow \top; p \rightarrow \mathbf{s} \}   \\
\mathbf{6}: \texttt{h(p)} & \{ \frac{1}{\sqrt{2}} (|000\rangle + |100\rangle)_{p,q,z}\} & \\
            &  \color{blue} \pi_\varsigma=\{\{p\},\{q,z\}\}  & \color{blue} \pi_\iota=\{\{p\},\{q,z\}\} &  \color{blue}  b=\{q, z \rightarrow \top; p \rightarrow \mathbf{d} \}   \\
\mathbf{7}: \texttt{cx(p,q)} & \{ \frac{1}{\sqrt{2}} (|000\rangle + |110\rangle)_{p,q,z}\} & \\
            &  \color{blue} \pi_\varsigma=\{\{p,q,z\}\}  & \color{blue} \pi_\iota=\{\{p\},\{q,z\}\} &  \color{blue}  b=\{p, q, z \rightarrow \top \}   \\
\mathbf{8}: \texttt{cx(z,q)} & \{ \frac{1}{\sqrt{2}} (|000\rangle + |110\rangle)_{p,q,z}\} & \\
            &  \color{blue} \pi_\varsigma=\{\{p,z\},\{q\}\}  & \color{blue} \pi_\iota=\{\{p\},\{q\},\{z\}\} &  \color{blue}  b=\{p, z \rightarrow \top; q \rightarrow \mathbf{s} \} 
\end{array}
\]
\caption{Illustration of the dangers of leveling two qubits when applying a CX gate with \textbf{$\top$} control qubit and \textbf{s} target qubit}
\label{xxx}
\end{table}

\subsection{Split and Join implementations}
\begin{verbatim}
fun join i arr =
    let 
        val (x1, x2) = (Array.sub(arr, i), Array.sub(arr, i+1))
        val (min_val, max_val) = (Int.min (x1, x2), Int.max (x1, x2))
    in 
        Array.modify (fn x => if x = max_val then min_val else x) arr
    end



fun split i arr =
    let val elm = Array.sub (arr, i)
    in 
        if i = elm then
            case Array.findi (fn (idx, x) => idx <> i andalso x = elm) arr of
                SOME (new_elm, _) => Array.modifyi (fn (idx, x) => 
                                        if idx <> i andalso x = elm then new_elm 
                                        else x) arr
                |NONE => ()
        else
            Array.update(arr, i, i)
    end
\end{verbatim}
\label{yyy}

\newpage
\subsection{Alternate Semantics} \label{sec:app2}

Above is presented an alternate set of analysis semantics, where qubits are represented as the abstract entities $q_i \in Q$ as opposed to integers. This version, although more verbose, is consistent with the concrete definition of our abstract domain, and its description throughout the report. As before, we use the symbol $i$ to denote the height/qubit indicies.

\begin{equation*}
\begin{split}
\llbracket \texttt{I} \rrbracket^\natural_i (b, \pi_\varsigma, \pi_\iota) &= (b, \pi_\varsigma, \pi_\iota) \\
\\
\llbracket C_1 \doubletimes C_2 \rrbracket^\natural_i (b, \pi_\varsigma, \pi_\iota) &= \llbracket C_2 \rrbracket^\natural_{i'} \circ \llbracket C_1 \rrbracket^\natural_i (b, \pi_\varsigma, \pi_\iota) \textit{ where } i' = i + \text{height}(C_1) \\
\\
\llbracket C_1 \doubleplus C_2 \rrbracket^\natural_i (b, \pi_\varsigma, \pi_\iota) &= \llbracket C_2 \rrbracket^\natural_i \circ \llbracket C_1 \rrbracket^\natural_i (b, \pi_\varsigma, \pi_\iota) \\
\\
\llbracket \sigma \rrbracket^\natural_i (b, \pi_\varsigma, \pi_\iota) &= (b, \pi_\varsigma, \pi_\iota) \\
\\
\llbracket \texttt{H} \rrbracket^\natural_i (b, \pi_\varsigma, \pi_\iota) &= (b^{q_i \mapsto \textbf{d}}, \pi_\varsigma, \pi_\iota) \textit{ if } b^{q_i} = \textbf{s} \\
&= (b^{q_i \mapsto \textbf{s}}, \pi_\varsigma, \pi_\iota) \textit{ if } b^{q_i} = \textbf{d} \\
&= (b, \pi_\varsigma, \pi_\iota \backslash q_i) \textit{ otherwise} \\
\\
\llbracket \texttt{T} \rrbracket^\natural_i (b, \pi_\varsigma, \pi_\iota) &= (b^{q_i \mapsto \top}, \pi_\varsigma, \pi_\iota) \textit{ if } b^{q_i} = \textbf{d} \\
&= (b, \pi_\varsigma, \pi_\iota) \textit{ otherwise} \\
\\
\llbracket \texttt{CX} \rrbracket^\natural_i (b, \pi_\varsigma, \pi_\iota) &= (b, \pi_\varsigma, \pi_\iota) \textit{ if } b^{q_i} = \mathbf{s} \textit{ or } b^{q_{(i+1)}} = \mathbf{d} \\
&= (b^{q_i, q_{(i+1)} \mapsto \top}, \pi_\varsigma \vee [q_i, q_{(i+1)}], \pi_\iota \vee [q_i, q_{(i+1)}]) \textit{ if } b^{q_i} = \mathbf{d} \textit{ and } b^{q_{(i+1)}} = \mathbf{s} \\
&= (b^{q_{(i+1)} \mapsto \mathbf{s}}, \pi_\varsigma \backslash q_{(i+1)}, \pi_\iota \backslash q_{(i+1)}) \textit{ if } \exists \mathcal{P} \in \pi_\iota, \textit{ where } \{q_i, q_{(i+1)}\} \subseteq \mathcal{P}  \\
&= (b^{q_i, q_{(i+1)} \mapsto \top}, \pi_\varsigma \vee [q_i, q_{(i+1)}], \pi_\iota) \; \textit{otherwise} \\
\\
\llbracket \texttt{SW} \rrbracket^\natural_{q_i} (b, \pi_\varsigma, \pi_\iota) &= (b, \pi_\varsigma, \pi_\iota)[q_i \mapsto q_{(i+1)}, q_{(i+1)} \mapsto q_i]
\end{split}
\end{equation*}

\end{document}